\documentclass[11pt]{article}

\usepackage[margin=1in]{geometry}
\usepackage[T1]{fontenc}
\usepackage[utf8]{inputenc}

\usepackage{amsmath,amssymb,mathtools}
\usepackage{graphicx}
\usepackage{booktabs}
\usepackage{url}
\usepackage{hyperref}
\usepackage{xcolor}
\usepackage{enumitem}
\usepackage{caption}

\hypersetup{
  colorlinks=true,
  linkcolor=blue,
  citecolor=blue,
  urlcolor=blue
}

\title{\textbf{3DAE: Binaural Quality Assessment for Audio Novel View Synthesis with Spatial Maps and Benchmark}}

\author{
\textbf{Jialu Xu}$^{*}$, \textbf{Yifan Zhou}$^{*}$\\
University of Waterloo\\
{\small $^{*}$Equal contribution}
}
\date{}

\begin{document}
\maketitle

\begin{abstract}
3D audio and novel-view acoustic synthesis models are usually evaluated with global metrics.
However, global metrics often hide where and why binaural prediction fails.
We propose a full-reference diagnostic framework that uses time-frequency audio error maps for magnitude, ILD, IPD, temporal alignment, loudness, and high-frequency failures, forming a 3D Audio Error Map (3DAE Map) for visual inspection.
We frame these diagnostics into a model-agnostic benchmark, Spatial Audio Error Bench (3DAE Bench), which takes arbitrary ground-truth and predicted binaural pairs and reports the prediction quality of audio novel-view synthesis models.
Experiments on ViGAS outputs over Replay-NVAS and SoundSpaces show different dominant failure modes: temporal misalignment on Replay-NVAS and ILD mismatch on SoundSpaces.
Overall, the framework provides interpretable failure-mode summaries and intuitive visual maps for audio Novel-view-synthesis model development optimization.
\end{abstract}


\section{Introduction}
Novel-view acoustic synthesis models aim to generate binaural audio at a new listening position~\cite{chen2023nvas}. Common evaluation metrics---including waveform error, STFT-based metrics, and reverberation measures~\cite{iso33821} may be useful for model ranking, but cannot properly reflect source of error. For instance, a model with low RMSE may still exhibit noticeable failures in localization cues, ILD, IPD, or timing~\cite{blauert1997spatial}.
We propose two components.
\begin{enumerate}
\item \textbf{3D Audio Error Maps}: An interactive 3D time-frequency diagnostic tool that exposes magnitude, ILD, IPD, and temporal/loudness errors of binaural prediction.
\item\textbf{3D Audio Error Benchmark}: A benchmark, given a ground-truth/predicted binaural pair from any model, reports a score vector, a dominant failure mode, and a benchmark report under a failure taxonomy spanning spectral error, ILD/IPD error, temporal misalignment, loudness mismatch, high-frequency loss, and data-quality warnings.
\end{enumerate}
We evaluate ViGAS~\cite{chen2023nvas} outputs on Replay-NVAS~\cite{shapovalov2023replay} and SoundSpaces~\cite{chen2022soundspaces2}. The same model exhibits different dominant failure modes across the two settings: temporal misalignment dominates on Replay-NVAS, while ILD mismatch dominates on SoundSpaces. This demonstrates that a single global score is insufficient for diagnosing binaural prediction failures.

\section{Related Work}
\subsection{Audio Novel-View Synthesis}
Audio novel-view synthesis renders the sound perceived at an unseen listening viewpoint from source observations. Recent work has introduced large-scale real and synthetic multi-view audio-visual datasets (e.g., Replay-NVAS~\cite{shapovalov2023replay}, SoundSpaces-NVAS~\cite{chen2022soundspaces2}) and visually guided synthesis models such as ViGAS~\cite{chen2023nvas}. More recent approaches explored scene-aware acoustic rendering via reconstructed geometry or neural fields~\cite{liang2023avnerf,brunetto2025neraf,bhosale2024avgs}. These efforts build on earlier work in audio-visual learning and binaural generation~\cite{gao2019visualsound,richard2021binaural,chen2020soundspaces}. However, evaluation in these systems is typically limited to global waveform, spectral, or acoustic summary metrics, optionally supplemented by qualitative spectrogram plots. Such summaries support model ranking but offer limited explanation of error origin---whether from temporal misalignment, binaural cue mismatch, loudness differences, or high-frequency loss. Our work provides an additional diagnostic layer for understanding how and why binaural predictions fail.
\subsection{Spatial Audio Evaluation Metrics}
Waveform and time-frequency-domain metrics---waveform error, STFT magnitude error, spectral distance, signal distortion---are widely used in audio generation and acoustic rendering. Physically motivated metrics such as reverberation time and clarity measures appear in room-acoustic analysis~\cite{iso33821}. For binaural audio, interaural cues (ILD, IPD, ITD) quantify perceived direction and spatial stability~\cite{blauert1997spatial}, with the associated head-related transfer functions (HRTFs) encoding these directional effects~\cite{moller1992binaural}. However, these metrics are typically applied in isolation; without integrating spatial, temporal, and acoustic diagnostics together, 3D rendering errors remain difficult to identify and interpret.
\subsection{3D Audio Visualization}
Unlike rendered images from 3D Gaussian splatting~\cite{kerbl2023gaussian}, audio differences are difficult to compare by ear alone. Subtle spatial errors can be hard to detect and track during model development. Spectrograms have long been used for qualitative audio inspection, but a plain spectrogram does not explicitly expose left/right ear errors, ILD mismatch, phase inconsistency, temporal offset, or loudness mismatch. We therefore propose an intuitive 3D audio error map that turns visual inspection into a debugging signal with interactive clicks, together with a benchmark interface for systematic analysis.

\section{Method}

\subsection{Setup}

Given a ground-truth binaural signal $x^{gt} \in \mathbb{R}^{T \times 2}$ and a predicted binaural signal $x^{pred} \in \mathbb{R}^{T \times 2}$, our goal is to diagnose where and how the prediction deviates from the reference. We assume a full-reference setting in which each prediction is paired with the corresponding ground truth. Inputs must be two-channel; mono signals cannot be used as either ground truth or prediction.

We treat the ground-truth sample rate as canonical and resample the prediction when needed, recording every resampling step in metadata so later errors can be traced back to preprocessing. This avoids fixing a global rate and keeps the framework portable across datasets and models. By default we adopt a strict duration policy: after sample-rate normalization, the two signals must contain the same number of samples. Nonetheless, equal length is necessary for pointwise comparison but not sufficient for temporal alignment, which we handle separately below. Each analysis produces a fixed set of outputs: metadata, validation warnings, global metrics, time-frequency maps, and summary statistics.

\subsection{Data Diagnostics and Temporal Alignment}

Before constructing error maps, we screen the inputs so that downstream maps are not driven by silence, clipping, level mismatch, or timing offset. Near-silent channels are flagged because ILD and IPD become unstable when one or both ears carry little energy. For each channel we compute
\begin{equation}
\mathrm{RMS}_c = \sqrt{\mathrm{mean}(x_c^2)},
\quad
\mathrm{silent\_ratio}_c = \mathrm{mean}(|x_c| < \tau_{silent}),
\end{equation}
and raise a near-silent warning when $\mathrm{RMS}_c < \tau_{RMS}$, with defaults $\tau_{RMS}=10^{-4}$ and $\tau_{silent}=10^{-5}$.

Clipping is detected via
\begin{equation}
\mathrm{clipping\_ratio} = \mathrm{mean}(|x| \geq \tau_{clip}), \quad \tau_{clip}=0.999,
\end{equation}
and treated as a data-quality warning rather than a spatial failure. We further compute the RMS ratio $r = \mathrm{RMS}_{pred} / \mathrm{RMS}_{gt}$, and warn when $r < 0.25$ or $r > 4$. Loudness mismatch is reported as its own failure mode rather than silently corrected, since level error can itself be a model failure and may otherwise dominate magnitude maps.

Equal sample counts do not imply that sound events line up in time. We estimate a global temporal offset using cross-correlation of stereo energy envelopes,
\begin{equation}
  e(t) = \mathrm{mean}(x_L(t)^2 + x_R(t)^2),
\qquad
  E[k] = \mathrm{mean}_{t \in \mathrm{frame}\ k} e(t),
\end{equation}
\begin{equation}
  d^* = \arg\max_d \mathrm{corr}\big(E^{gt}, \mathrm{shift}(E^{pred}, d)\big),
\end{equation}
restricting the search to $\pm 100$ ms. A positive $d^*$ means the prediction is late relative to ground truth. We raise a temporal warning when $|d^*| \geq 5$ ms, a strong warning when $|d^*| \geq 20$ ms, and mark the estimate as low-confidence when the correlation peak falls below $0.3$. We do not realign the waveform, because temporal mismatch is itself part of the prediction error; the warnings are instead propagated so that magnitude, ILD, and IPD maps are read in their context.

\subsection{Time-Frequency Error Maps}

After validation, we transform both signals into a shared time-frequency representation. For each channel $c \in \{L,R\}$ we compute
\begin{equation}
X^{gt}_c(f,n) = \mathrm{STFT}(x^{gt}_c), \quad
X^{pred}_c(f,n) = \mathrm{STFT}(x^{pred}_c),
\end{equation}
using a Hann window with a 32 ms window length and an 8 ms hop~\cite{oppenheim2009dsp}. Window and hop are specified in milliseconds and converted to samples using the ground-truth rate, and the frequency/time axes are saved with each output for reproducibility.

\textbf{Magnitude.} The first map measures log-magnitude error per ear,
\begin{equation}
E^{mag}_c(f,n) = \big| \log(|X^{gt}_c| + \epsilon) - \log(|X^{pred}_c| + \epsilon) \big|,
\end{equation}
together with its stereo mean $E^{mag}_{mean} = (E^{mag}_L + E^{mag}_R)/2$. We keep both ear-specific and stereo-mean maps because asymmetric, one-sided artifacts would be averaged out otherwise.

\textbf{ILD.} To diagnose level cues, we form a per-ear log-magnitude difference and take its absolute error,
\begin{equation}
ILD^{*}(f,n) = \log(|X^{*}_L| + \epsilon) - \log(|X^{*}_R| + \epsilon),
\quad * \in \{gt, pred\},
\end{equation}
\begin{equation}
E^{ILD}(f,n) = |ILD^{gt}(f,n) - ILD^{pred}(f,n)|.
\end{equation}
ILD is unreliable in low-energy regions, so we mask bins by ground-truth stereo magnitude $A^{gt} = (|X^{gt}_L| + |X^{gt}_R|)/2$ and keep a bin only when $A^{gt}$ exceeds the configured energy threshold. Invalid bins are stored explicitly as masked values rather than zero error, and ILD summaries are reported only over valid bins together with the valid-bin ratio. This distinguishes a small ILD error from an absence of reliable ILD evidence.

\textbf{IPD.} Phase cues are measured with the circular phase distance,
\begin{equation}
IPD^{*}(f,n) = \angle X^{*}_L(f,n) - \angle X^{*}_R(f,n),
\quad * \in \{gt, pred\},
\end{equation}
\begin{equation}
E^{IPD}(f,n) = \big| \angle \exp\!\big(i\, (IPD^{gt} - IPD^{pred})\big) \big|.
\end{equation}
The same energy mask as ILD is applied, and the default interpretation is restricted to frequencies below 1.5 kHz; above this range, phase wrapping and small timing offsets can dominate the IPD difference and make it unstable~\cite{blauert1997spatial}.

For every map we store the raw array and a heatmap visualization, and report mean, median, 95th percentile, maximum, and valid-bin ratio where applicable. We additionally compute a high-frequency magnitude diagnostic---the stereo magnitude error averaged above 4 kHz---which captures cases where the prediction sounds muffled or lacks high-frequency detail. All maps are computed from the validated GT/pred pair without temporal or loudness correction. This preserves the model outputs' inherent problems, but it also means that map readings must be interpreted alongside the earlier warnings: severe loudness mismatch can inflate magnitude maps, and a strong temporal offset can inflate spectral and high-frequency errors that are not, in themselves, spectral modeling failures.

\subsection{Benchmark: Failure-Mode Scores}

We collapse each map into scalar summaries (mean, median, 95th percentile, max, and valid-bin ratio where applicable) and organize them into a failure-mode vector:
\begin{itemize}
  \item \texttt{spectral\_error\_score}: stereo-mean log-magnitude error.
  \item \texttt{ear\_specific\_error\_score}: left/right magnitude asymmetry.
  \item \texttt{ild\_error\_score}: masked ILD error.
  \item \texttt{ipd\_error\_score}: masked low/mid-frequency circular IPD error.
  \item \texttt{temporal\_misalignment\_score}: estimated global delay.
  \item \texttt{loudness\_mismatch\_score}: predicted-to-ground-truth RMS ratio.
  \item \texttt{high\_frequency\_loss\_score}: magnitude error above 4 kHz.
  \item \texttt{data\_quality\_warning\_score}: clipping, silence, and validation warnings.
\end{itemize}
Several scores have closed-form normalizations:
\begin{equation}
\texttt{temporal\_misalignment\_score} = |\mathrm{delay}_{ms}| / 20,
\end{equation}
\begin{equation}
\texttt{loudness\_mismatch\_score} = |\log(r)| / \log(4),
\end{equation}
\begin{equation}
\texttt{ipd\_error\_score} = \mathrm{mean}(E^{IPD}) / \pi.
\end{equation}

The raw dominant failure mode is $\arg\max$ over the score vector. To prevent magnitude-related scores from being inflated by errors that originate from gross timing or level mismatch, we assign the reported dominant mode under warning-aware logic:
\begin{equation}
\begin{aligned}
&\text{if } |\mathrm{delay}_{ms}| \geq 20, &&\texttt{main} = \texttt{temporal\_misalignment};\\
&\text{else if } r < 0.25 \text{ or } r > 4, &&\texttt{main} = \texttt{loudness\_mismatch};\\
&\text{otherwise}, &&\texttt{main} = \texttt{raw\_main}.
\end{aligned}
\end{equation}
All raw scores are still reported to reflect output's inherent problems; only the headline interpretation is adjusted.

\subsection{Benchmark Design}

The benchmark is a full-reference test environment that is generally compatible to dataset and model: the only requirement is that a model can provide paired ground-truth and predicted binaural waveforms. Each evaluation run is specified by a minimal CSV manifest,
\begin{verbatim}
pair_id,gt_audio_path,pred_audio_path
\end{verbatim}
so predictions from ViGAS~\cite{chen2023nvas}, NeRAF-style reconstruction pipelines~\cite{brunetto2025neraf,mildenhall2020nerf}, or any future audio NVS model can be evaluated uniformly.

Each run begins with audio preflight checks for missing or unreadable files, unsupported formats, non-stereo audio, invalid duration, and sample-rate mismatch requiring resampling. Invalid pairs are skipped and recorded in \texttt{failed\_pairs.csv} so that a large run is not interrupted by a small number of broken samples. Valid pairs then go through validation, temporal diagnostics, map computation, score extraction, and warning-aware interpretation.

\begin{figure}[t]
  \centering
  \includegraphics[width=0.98\linewidth]{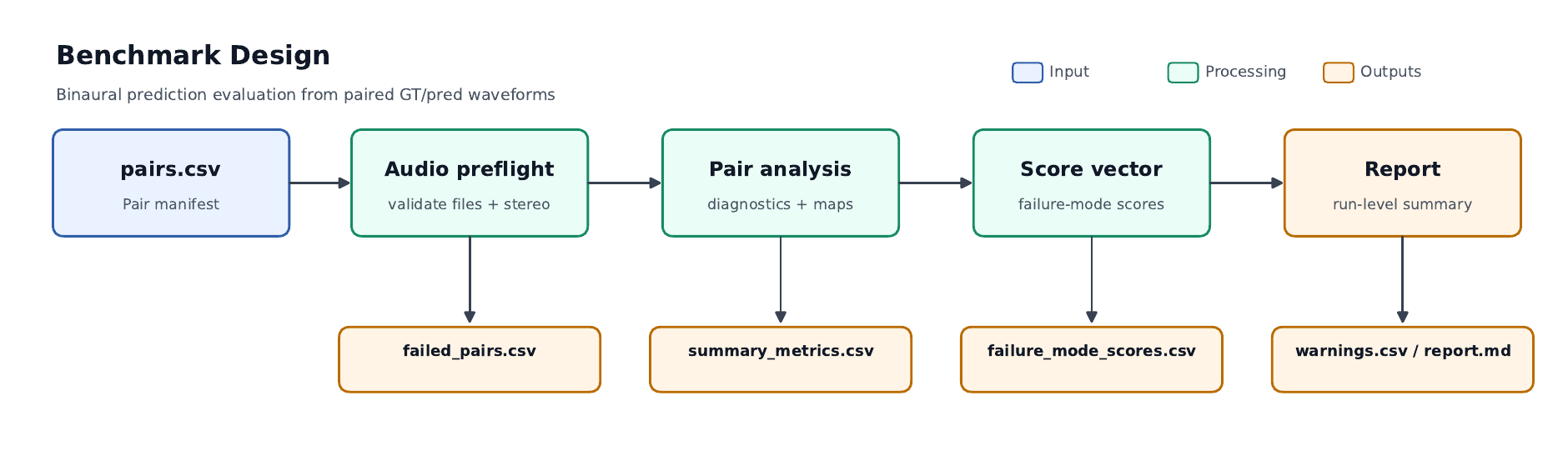}
  \caption{Benchmark design.}
  \label{fig:benchmark_pipeline}
\end{figure}

The main outputs are \texttt{summary\_metrics.csv}, \texttt{failure\_mode\_scores.csv}, \texttt{benchmark\_scores.json}, \texttt{warnings.csv}, \texttt{failed\_pairs.csv}, and \texttt{report.md}. The CSV files preserve pair-level raw metrics, failure-mode scores, warnings, and failed samples; the JSON and Markdown report summarize the run-level score vector, dominant failure mode, warning counts, and worst cases. Pair-level vectors are averaged into a run-level summary, and the run-level dominant failure mode follows the same warning-aware logic: when many pairs carry strong temporal or loudness warnings, the spectral and high-frequency means are not, by default, attributed to spectral modeling failures. For large experiments, a summary-only mode retains aggregate metrics and reports while discarding per-pair raw maps and figures after score extraction, which keeps full Replay-NVAS and SoundSpaces evaluations practical.

\section{Experiments}

\subsection{Setup}

We evaluate the benchmark on two binaural NVS datasets with ViGAS~\cite{chen2023nvas} outputs: real-scene Replay-NVAS~\cite{shapovalov2023replay} ($N{=}233$ pairs from the test split) and synthetic SoundSpaces-NVAS~\cite{chen2022soundspaces2} ($N{=}426$ pairs). The benchmark runs in summary-only mode---per-pair metrics The run-level report are obtained, raw maps and figures are released only for representative cases. Any alignment or loudness correction are not applied. Timing and level mismatches remain visible to be failure modes.

\subsection{Controlled distortion}

A suite of controlled distortions of Replay-NVAS ground-truth audio were used to verify design. Identity yields zero error on all maps; global $-6$ dB raises per-ear magnitude while leaving ILD and IPD near zero; one-ear $-6$ dB produces a clean ear-specific asymmetry and a matching ILD error; mono-copy and channel swap leave per-ear magnitude intact but produce largest ILD errors in the suite. Another case is the $1$ ms right-channel delay: it does not perturb the stereo energy envelope and raises no global temporal warning, but IPD map recorded a circular phase error of $\sim\!1.56$ rad. The temporal envelope and the IPD map are complementary in recording global offsets and sub-frame interaural shifts.

\subsection{Replay-NVAS ViGAS Benchmark}

The run-level mean score vector is dominated by temporal misalignment ($1.39$), followed by spectral error ($0.86$), high-frequency loss ($0.85$), ILD ($0.67$), and IPD ($0.40$); ear-specific asymmetry remains low ($0.05$). Of the $233$ pairs, $94$ trigger a strong temporal warning ($|d^*|\!\geq\!20$ ms) and $208$ carry at least one validation warning, so the warning-aware rule assigns the run-level dominant mode to temporal misalignment rather than to the spectral or high-frequency maps.

A waveform RMSE or magnitude-only readout may record this run as spectral issue without exposing the timing cause. Since the spectral and high-frequency maps are influences by the global timing offset, the warning-aware label reattributes the dominant mode to the underlying cause while preserving every raw score for inspection. The pair-level counts---temporal dominant for $119/233$ pairs, spectral for another $86$---back the same interpretation. A potential reason may attribute to such phenomena: Replay-NVAS captures the source and target views with physical microphones at different positions. The onset alignment between viewpoints may not be perfect, and models that do not explicitly handle this offset will accumulate temporal error.

\subsection{SoundSpaces ViGAS Benchmark}

The score vector for this test is dominated by ILD mismatch ($0.69$), followed by high-frequency loss ($0.67$) and spectral error ($0.65$); temporal misalignment drops to $0.17$ and only $29$ pairs trigger a strong temporal warning. The dominant failure here is the left/right level relationship.

This represents another failure mode that traditional metrics may obscure. When evaluated independently for each ear, the predictions may appear accurate; however, primary error lies in the \textit{interaural difference}, which can be diluted or averaged out by stereo-mean magnitude errors. The ILD map can be used to handle such discrepancy. At the pair level, $298/426$ pairs are classified as ILD-dominant, whereas high-frequency loss is a much less frequent secondary failure mode ($65$ pairs). SoundSpaces is based on synthetic impulse responses and does not contain recording-induced temporal offsets. However, synthesized binaural cues place greater stress on the model's ability to generalize spatially, with ILD emerging as the first cue to degrade. Thus, the same model exhibits different dominant weaknesses across datasets. The score vector may be used to identify which perceptual or spatial cue is responsible for the degradation in each setting.

\begin{figure}[t]
  \centering
  \includegraphics[width=0.98\linewidth]{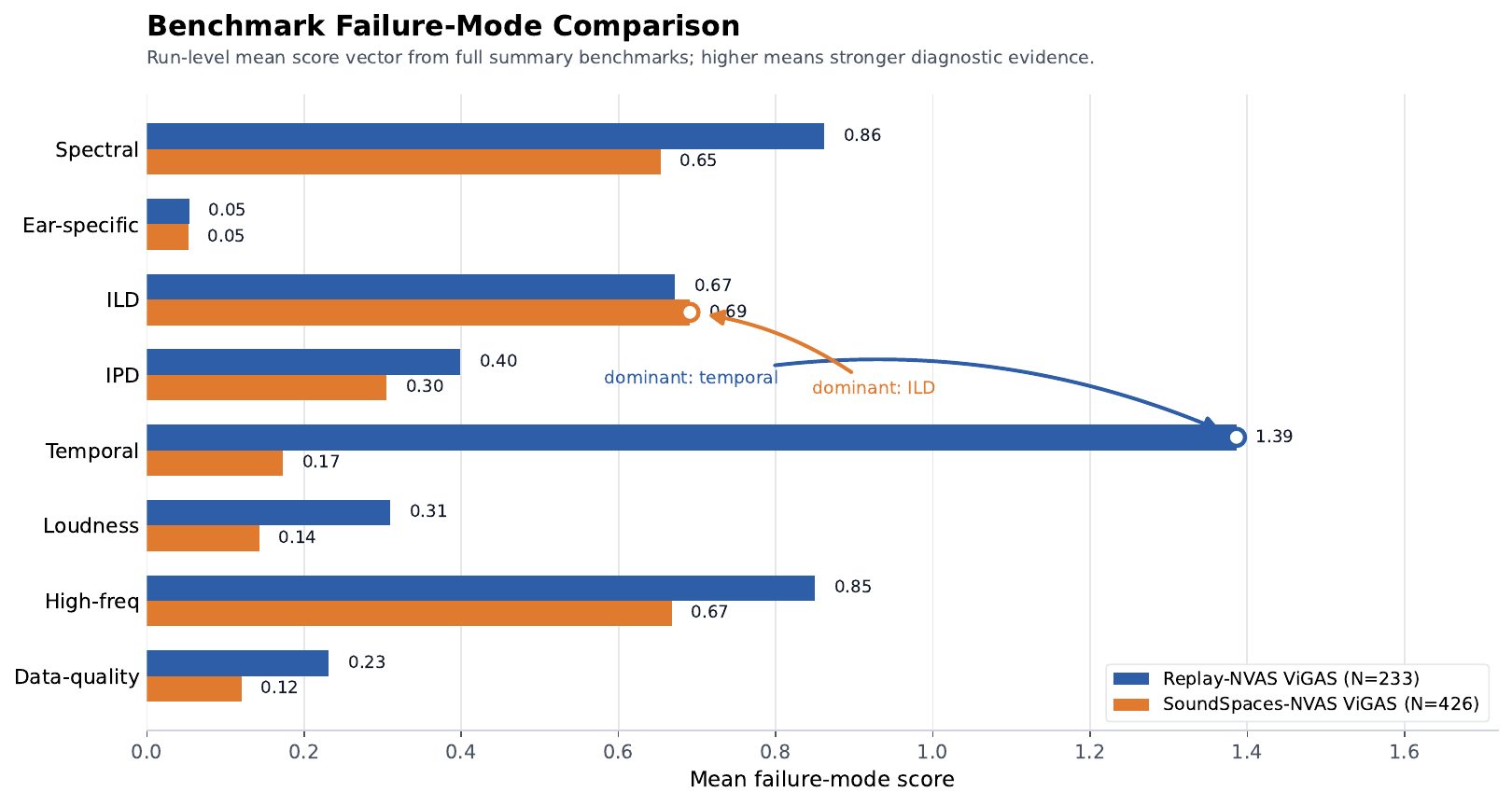}
  \caption{Benchmark failure-mode comparison for ViGAS on Replay-NVAS versus SoundSpaces-NVAS.}
  \label{fig:bench_compare}
\end{figure}

\begin{figure}[p]
  \centering
  \includegraphics[width=0.98\linewidth]{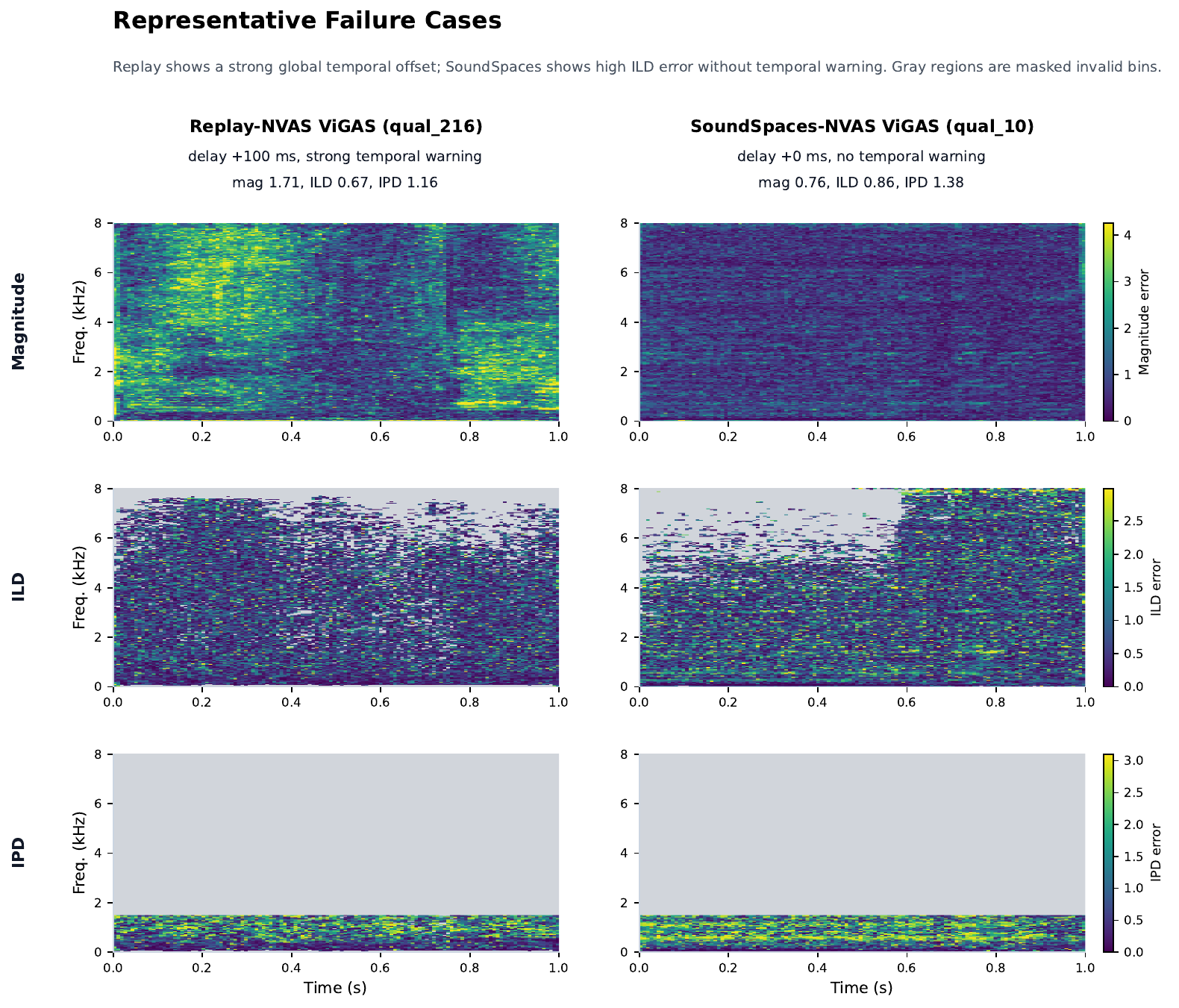}
  \caption{Representative failure cases: a temporal-dominant Replay-NVAS pair and an ILD-dominant SoundSpaces-NVAS pair, with their magnitude, ILD, and IPD maps.}
  \label{fig:bench_cases}
\end{figure}

\subsection{Discussion}

Three observations follow from these runs, all highlighting limitations of global metrics.

\textit{No single global metric identifies the failure mode.} Waveform RMSE, STFT magnitude error, and similar scalar summaries collapse spectral, binaural, temporal, and loudness errors into a single value. They can rank outputs by aggregate distortion, but they do not indicate which perceptual cue is responsible; this limitation is also recognized in subjective evaluation standards~\cite{itu1534}. In contrast, the score vector separates these cues and links each score to its corresponding diagnostic map. The Replay-NVAS/SoundSpaces-NVAS comparison illustrates this distinction: a scalar metric would label both runs as degraded, whereas the score vector specifies the dominant cue driving the degradation in each case.

\textit{The same model fails differently across datasets.} ViGAS~\cite{chen2023nvas} is primarily affected by temporal misalignment on Replay-NVAS~\cite{shapovalov2023replay}, but by ILD mismatch on SoundSpaces-NVAS~\cite{chen2022soundspaces2}. These two cases require different improvement strategies, although mean magnitude error or waveform RMSE would largely treat them as comparable failures. The proposed framework exposes this dataset-dependent shift without modifying the model, its outputs, or the evaluation inputs.

\textit{Warning-aware interpretation reduces misattribution.} Large timing or loudness errors can artificially increase magnitude-based and high-frequency scores. On Replay-NVAS, for example, the raw spectral score reaches $0.86$, but much of this increase is induced by $|d^*|\geq 20$ ms offsets in $94$ pairs. A raw $\arg\max$ rule would therefore attribute the failure to the spectral map, even when the underlying cause is temporal. The warning-aware rule preserves all raw scores but assigns the final interpretation to the causal cue. This produces a more defensible failure-mode label than simply selecting the largest scalar value.

\section{Conclusion}

We presented a full-reference 3D audio error framework for binaural prediction error visualization and package it into a benchmark system. From any pair of ground-truth and predicted binaural waveforms, the framework outputs per-ear and stereo magnitude maps, masked ILD and low/mid-frequency IPD maps, temporal-alignment, loudness, and data-quality diagnostics, and a failure-mode score vector with a warning-aware dominant mode and a per-run report. Applied to ViGAS, the same model is dominantly temporal-misaligned on Replay-NVAS and dominantly ILD-mismatched on SoundSpaces-NVAS. The framework can reveal binural rendering errors and dataset-depent issues that no single global metric exposes.

Our future work includes a human-aligned perceptual metric trained on subjective listening responses~\cite{itu1534,kumar2023squim} to validate and re-weight the current failure-mode scores for unified audio NVS model benchmarking.

\bibliographystyle{ieeetr}
\bibliography{references}

@inproceedings{chen2023nvas,
  author    = {Changan Chen and Alexander Richard and Roman Shapovalov and
               Vamsi Krishna Ithapu and Natalia Neverova and
               Kristen Grauman and Andrea Vedaldi},
  title     = {Novel-View Acoustic Synthesis},
  booktitle = {Proceedings of the IEEE/CVF Conference on Computer Vision
               and Pattern Recognition (CVPR)},
  pages     = {6409--6419},
  year      = {2023}
}

@inproceedings{shapovalov2023replay,
  author    = {Roman Shapovalov and Yanir Kleiman and Ignacio Rocco and
               David Novotny and Andrea Vedaldi and Changan Chen and
               Filippos Kokkinos and Ben Graham and Natalia Neverova},
  title     = {Replay: Multi-modal Multi-view Acted Videos for Casual Holography},
  booktitle = {Proceedings of the IEEE/CVF International Conference on
               Computer Vision (ICCV)},
  pages     = {20338--20348},
  year      = {2023}
}

@inproceedings{chen2022soundspaces2,
  author    = {Changan Chen and Carl Schissler and Sanchit Garg and
               Philip Kobernik and Alexander Clegg and Paul Calamia and
               Dhruv Batra and Philip W. Robinson and Kristen Grauman},
  title     = {{SoundSpaces} 2.0: A Simulation Platform for Visual-Acoustic Learning},
  booktitle = {Advances in Neural Information Processing Systems (NeurIPS),
               Datasets and Benchmarks Track},
  year      = {2022}
}

@inproceedings{liang2023avnerf,
  author    = {Susan Liang and Chao Huang and Yapeng Tian and
               Anurag Kumar and Chenliang Xu},
  title     = {{AV-NeRF}: Learning Neural Fields for Real-World Audio-Visual
               Scene Synthesis},
  booktitle = {Advances in Neural Information Processing Systems (NeurIPS)},
  year      = {2023}
}

@inproceedings{brunetto2025neraf,
  author    = {Amandine Brunetto and Sascha Hornauer and Fabien Moutarde},
  title     = {{NeRAF}: 3D Scene Infused Neural Radiance and Acoustic Fields},
  booktitle = {Proceedings of the International Conference on Learning
               Representations (ICLR)},
  year      = {2025},
  url       = {https://openreview.net/forum?id=njvSBvtiwp}
}

@inproceedings{bhosale2024avgs,
  author    = {Swapnil Bhosale and Haosen Yang and Diptesh Kanojia and
               Jiankang Deng and Xiatian Zhu},
  title     = {{AV-GS}: Learning Material and Geometry Aware Priors for
               Novel View Acoustic Synthesis},
  booktitle = {Advances in Neural Information Processing Systems (NeurIPS)},
  year      = {2024}
}

@book{blauert1997spatial,
  author    = {Jens Blauert},
  title     = {Spatial Hearing: The Psychophysics of Human Sound Localization},
  edition   = {Revised},
  publisher = {MIT Press},
  address   = {Cambridge, MA},
  year      = {1997},
  isbn      = {978-0-262-02413-6}
}

@techreport{iso33821,
  author      = {{International Organization for Standardization}},
  title       = {{ISO} 3382-1: Acoustics --- Measurement of Room Acoustic
                 Parameters --- Part~1: Performance Spaces},
  institution = {International Organization for Standardization},
  type        = {Standard},
  number      = {ISO 3382-1:2009},
  year        = {2009}
}

@inproceedings{chen2020soundspaces,
  author    = {Changan Chen and Unnat Jain and Carl Schissler and
               Sebastia Vicenc Amengual Gari and Ziad Al-Halah and
               Vamsi Krishna Ithapu and Philip Robinson and Kristen Grauman},
  title     = {{SoundSpaces}: Audio-Visual Navigation in {3D} Environments},
  booktitle = {Proceedings of the European Conference on Computer Vision (ECCV)},
  year      = {2020}
}

@inproceedings{gao2019visualsound,
  author    = {Ruohan Gao and Kristen Grauman},
  title     = {2.5{D} Visual Sound},
  booktitle = {Proceedings of the IEEE/CVF Conference on Computer Vision
               and Pattern Recognition (CVPR)},
  pages     = {324--333},
  year      = {2019}
}

@inproceedings{mildenhall2020nerf,
  author    = {Ben Mildenhall and Pratul P. Srinivasan and Matthew Tancik and
               Jonathan T. Barron and Ravi Ramamoorthi and Ren Ng},
  title     = {{NeRF}: Representing Scenes as Neural Radiance Fields for
               View Synthesis},
  booktitle = {Proceedings of the European Conference on Computer Vision (ECCV)},
  pages     = {405--421},
  year      = {2020}
}

@article{kerbl2023gaussian,
  author    = {Bernhard Kerbl and Georgios Kopanas and
               Thomas Leimk{\"u}hler and George Drettakis},
  title     = {3{D} Gaussian Splatting for Real-Time Radiance Field Rendering},
  journal   = {ACM Transactions on Graphics},
  volume    = {42},
  number    = {4},
  articleno = {139},
  pages     = {139:1--139:14},
  year      = {2023}
}

@techreport{itu1534,
  author      = {{International Telecommunication Union}},
  title       = {Method for the Subjective Assessment of Intermediate Quality
                 Level of Audio Systems ({MUSHRA})},
  institution = {International Telecommunication Union},
  type        = {Recommendation},
  number      = {ITU-R BS.1534-3},
  year        = {2015}
}

@inproceedings{richard2021binaural,
  author    = {Alexander Richard and Dejan Markovic and Israel D. Gebru and
               Steven Krenn and Gladstone Butler and Fernando de la Torre and
               Yaser Sheikh},
  title     = {Neural Synthesis of Binaural Speech from Mono Audio},
  booktitle = {International Conference on Learning Representations (ICLR)},
  year      = {2021}
}

@book{oppenheim2009dsp,
  author    = {Alan V. Oppenheim and Ronald W. Schafer},
  title     = {Discrete-Time Signal Processing},
  edition   = {Third},
  publisher = {Pearson Prentice Hall},
  year      = {2009},
  isbn      = {978-0-13-198842-2}
}

@article{moller1992binaural,
  author  = {Henrik M{\o}ller},
  title   = {Fundamentals of Binaural Technology},
  journal = {Applied Acoustics},
  volume  = {36},
  number  = {3--4},
  pages   = {171--218},
  year    = {1992}
}

@inproceedings{kumar2023squim,
  author    = {Anurag Kumar and Ke Tan and Zhaoheng Ni and Pranay Manocha and
               Xiaohui Zhang and Ethan Henderson and Buye Xu},
  title     = {{TorchAudio-Squim}: Reference-Less Speech Quality and
               Intelligibility Measures in {TorchAudio}},
  booktitle = {Proceedings of the IEEE International Conference on Acoustics,
               Speech and Signal Processing (ICASSP)},
  pages     = {1--5},
  year      = {2023}
}

\clearpage
\appendix
\section{3D Audio Error Maps}
\label{app:audio_visual_maps}

This appendix documents the visual interface used to inspect the proposed 3D Audio Error Maps.
The interface is intended for qualitative diagnosis after the quantitative maps and benchmark scores have been computed.
It supports three operation modes: direct analysis of an uploaded binaural pair, inspection of representative model examples, and scene-level exploration of spatial error patterns.
In all modes, brighter yellow colors denote larger local error values, while darker purple colors denote smaller errors.
Gray regions indicate bins that are masked as invalid or not strongly interpretable under the corresponding error-map definition.

\subsection{Mode 1: Direct Analysis of a Ground-Truth/Prediction Pair}
\label{app:mode1}

In the first mode, the user provides a ground-truth binaural waveform and a predicted binaural waveform.
The pair must be stereo; mono audio is rejected rather than silently converted.
After the two files are selected, clicking \texttt{Run Analysis} launches the same validation, temporal diagnostic, STFT analysis, and error-map computation used by the benchmark.
The left panel displays the ground-truth and predicted audio players, warnings, downloadable metadata, and the main summary metrics.
The map panels display stereo-mean magnitude error, individual left/right-ear magnitude errors, ILD error, and IPD error.
This mode is useful for inspecting a new model output without constructing a full benchmark manifest.

\begin{figure}[p]
  \centering
  \includegraphics[width=0.98\linewidth]{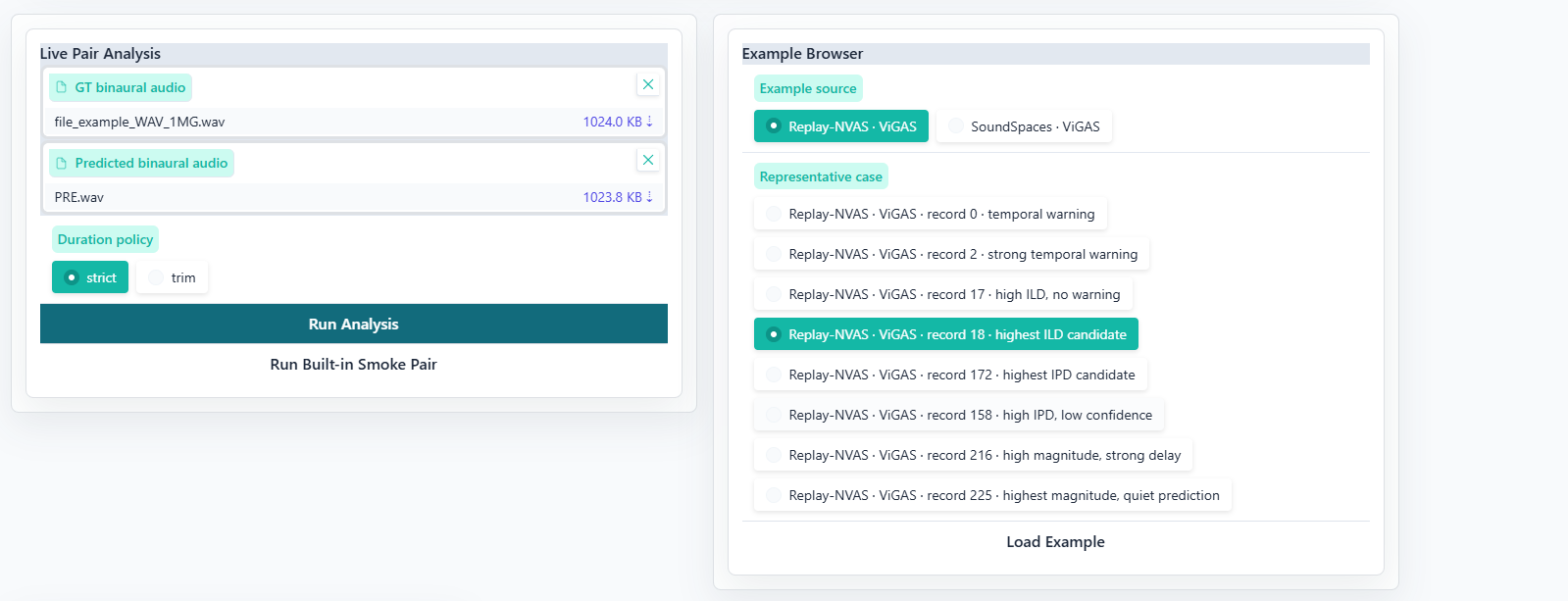}
  \vspace{0.6em}
  \includegraphics[width=0.98\linewidth]{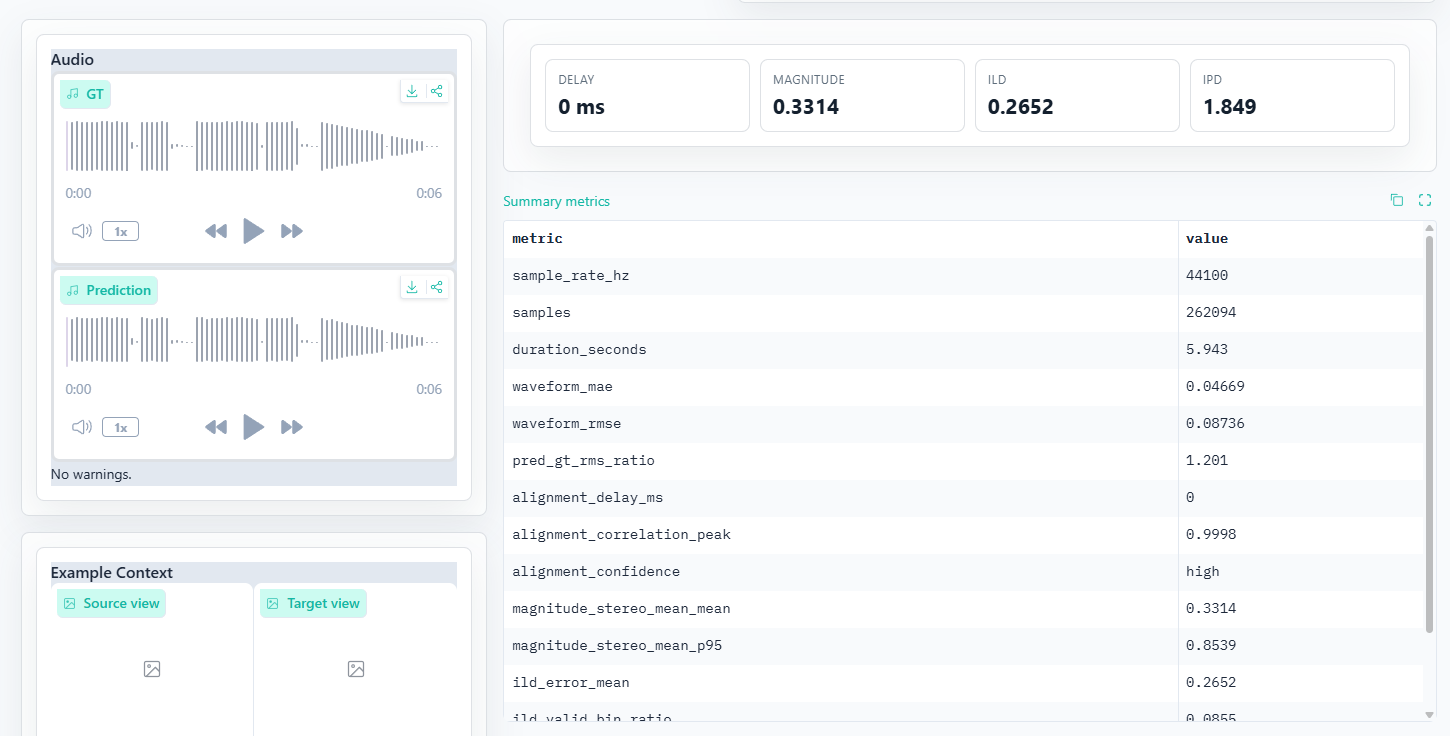}
  \caption{Mode 1 interface for direct analysis. The user uploads a ground-truth/prediction binaural pair and starts analysis with \texttt{Run Analysis}. The interface reports audio playback, warnings, metadata, and summary metrics before showing the corresponding error maps.}
  \label{fig:app_mode1_overview}
\end{figure}

\begin{figure}[p]
  \centering
  \includegraphics[width=0.74\linewidth]{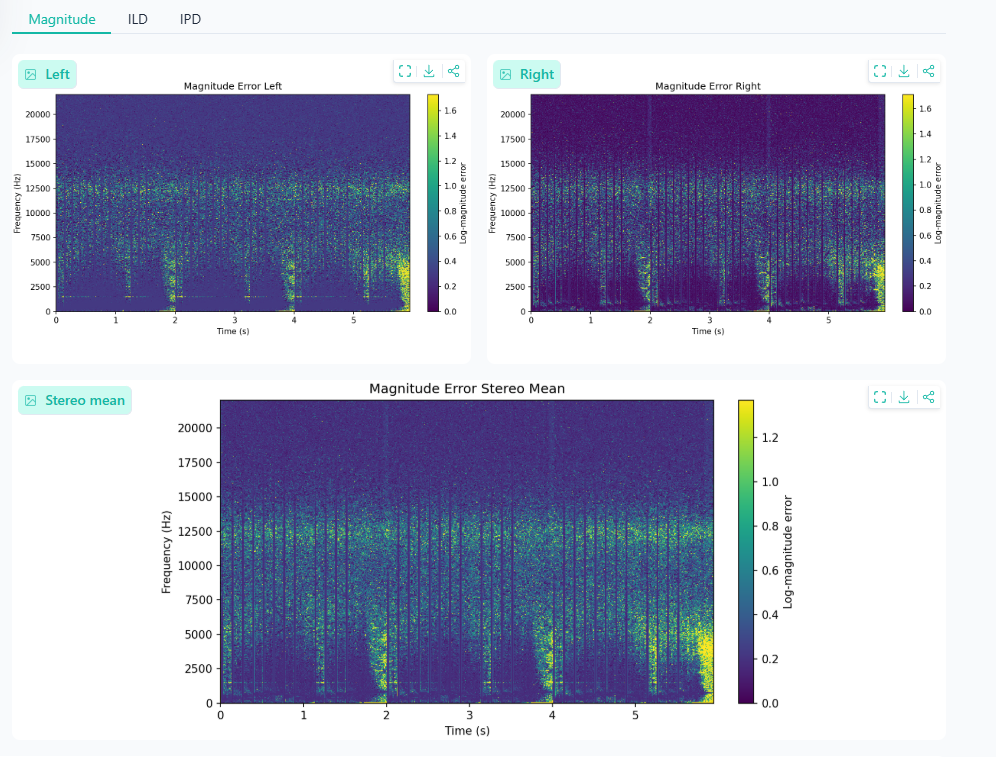}
  \vspace{0.6em}

  \begin{minipage}{0.49\linewidth}
    \centering
    \includegraphics[width=\linewidth]{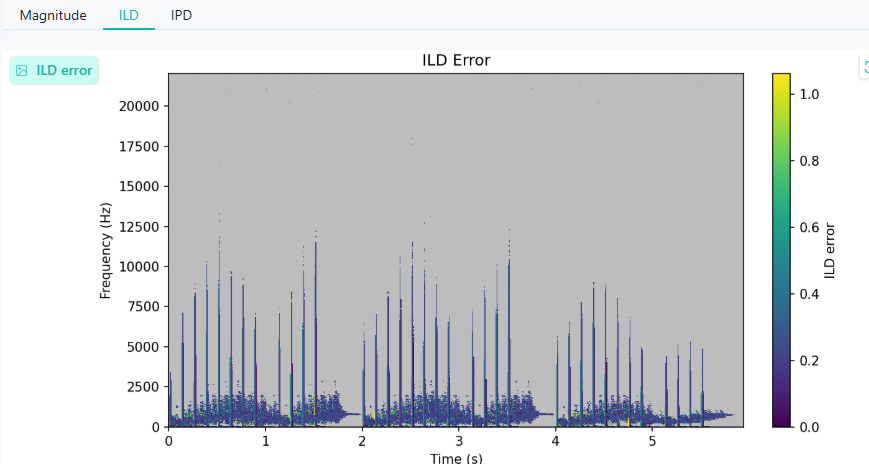}
  \end{minipage}
  \hfill
  \begin{minipage}{0.49\linewidth}
    \centering
    \includegraphics[width=\linewidth]{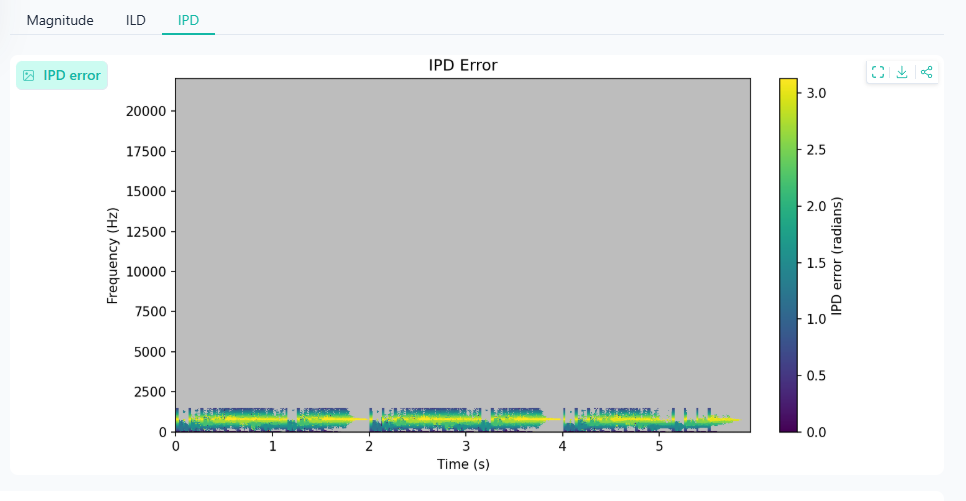}
  \end{minipage}
  \caption{Mode 1 error-map views. The interface shows stereo-mean magnitude error, individual ear-specific magnitude error, ILD error, and IPD error. These maps expose different failure types rather than reducing the pair to one global score.}
  \label{fig:app_mode1_maps}
\end{figure}

\subsection{Mode 2: Representative Dataset--Model Examples}
\label{app:mode2}

The second mode loads precomputed representative examples from dataset--model evaluations.
The user first chooses an example source, then selects a representative case, and finally clicks \texttt{Load Example}.
The loaded case contains the original ground-truth and predicted audio, analysis metadata, warning messages, summary scores, and the precomputed error maps.
This mode is designed for demonstration and comparison: it allows the same diagnostic interface to show temporal-dominant, ILD-dominant, high-frequency, or other selected failure cases without rerunning analysis live.

\begin{figure}[p]
  \centering
  \includegraphics[width=0.98\linewidth]{appendix_fig/1.png}
  \vspace{0.6em}
  \includegraphics[width=0.98\linewidth]{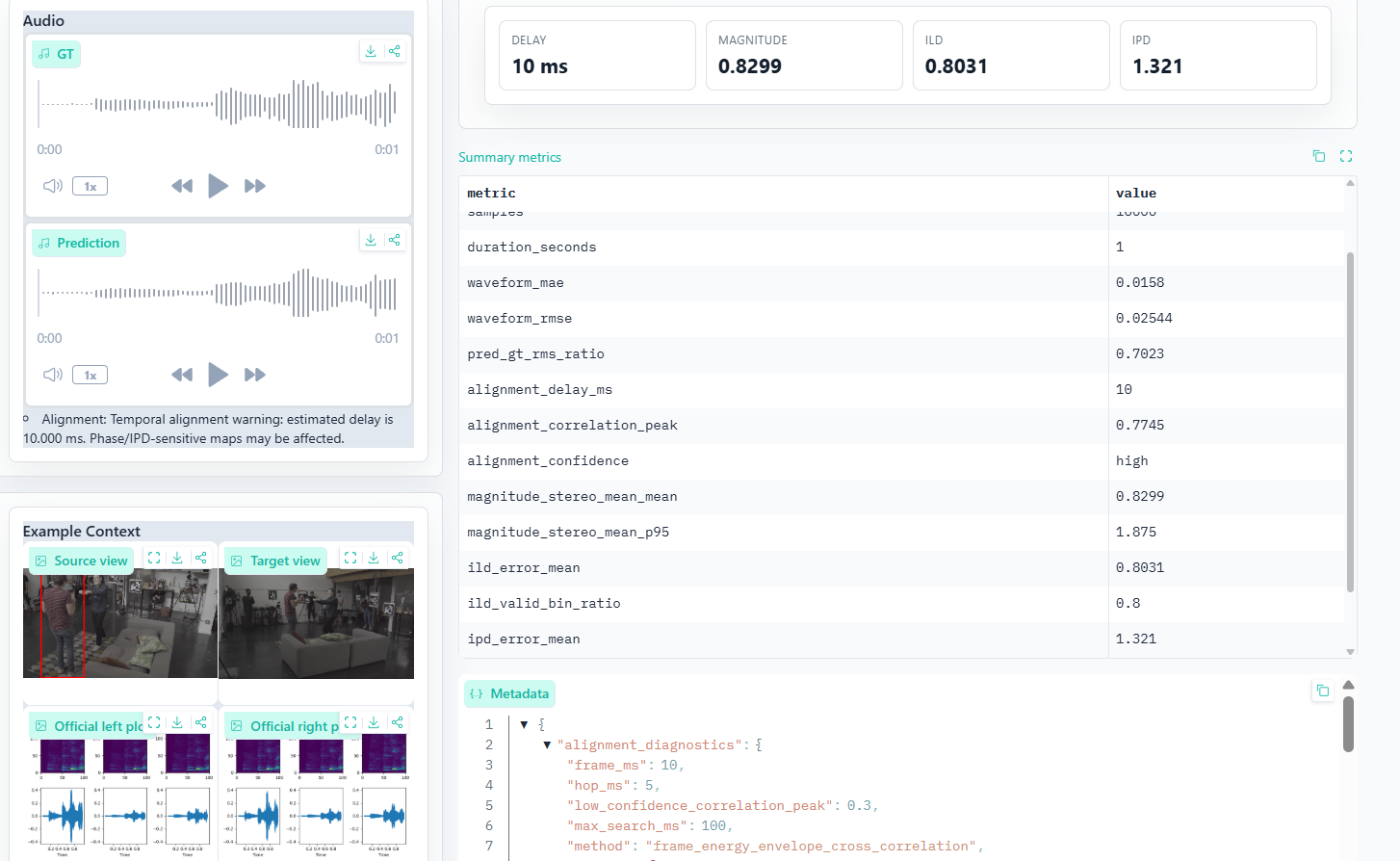}
  \caption{Mode 2 example browser. The user selects a dataset--model example source, chooses a representative case, and loads the precomputed analysis.}
  \label{fig:app_mode2_browser}
\end{figure}

\begin{figure}[p]
  \centering
  \includegraphics[width=0.74\linewidth]{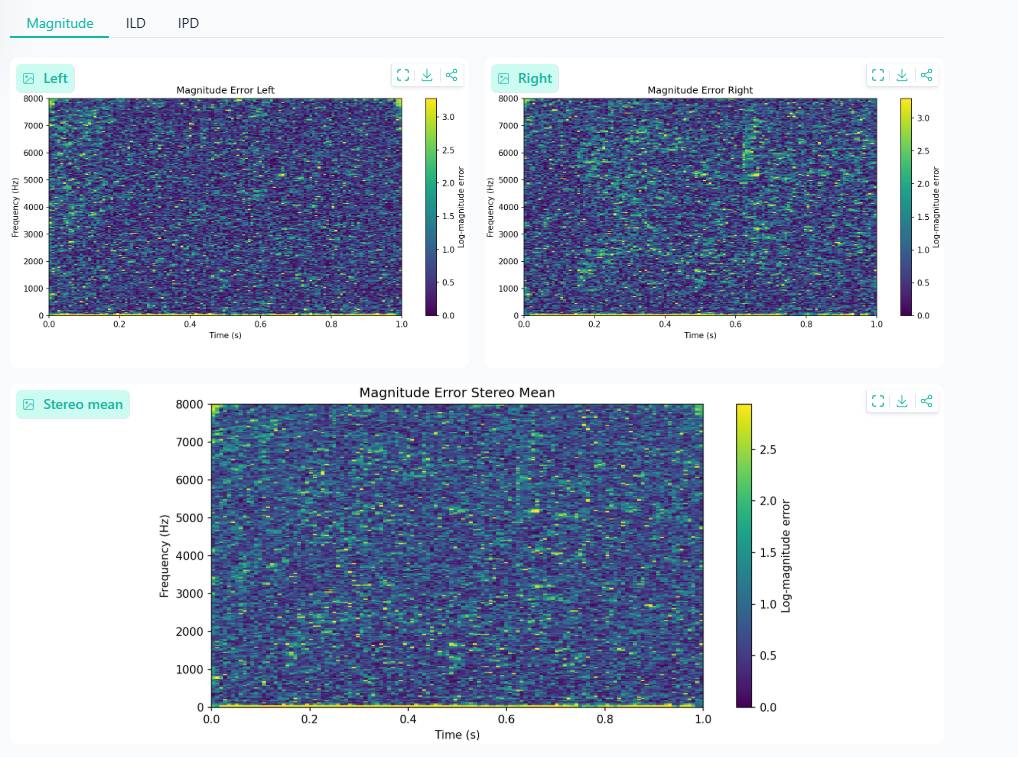}
  \vspace{0.6em}

  \begin{minipage}{0.49\linewidth}
    \centering
    \includegraphics[width=\linewidth]{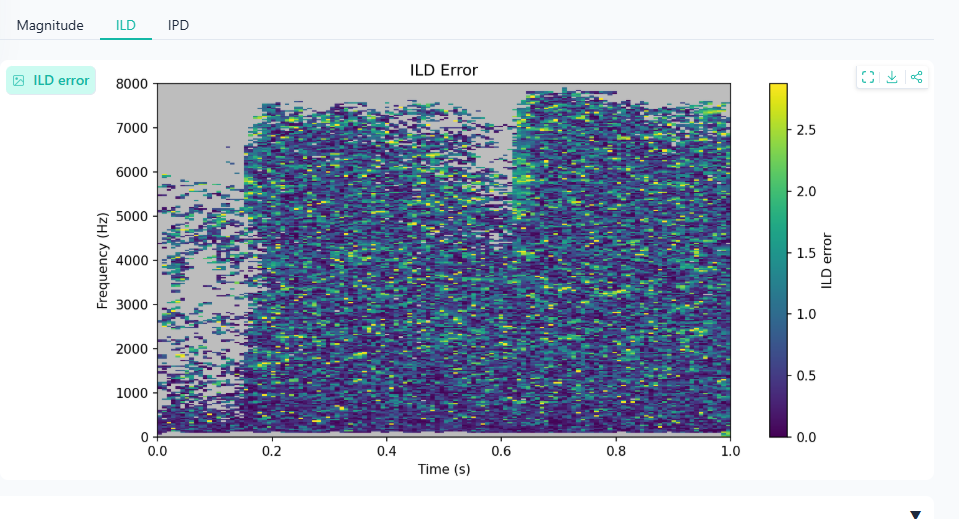}
  \end{minipage}
  \hfill
  \begin{minipage}{0.49\linewidth}
    \centering
    \includegraphics[width=\linewidth]{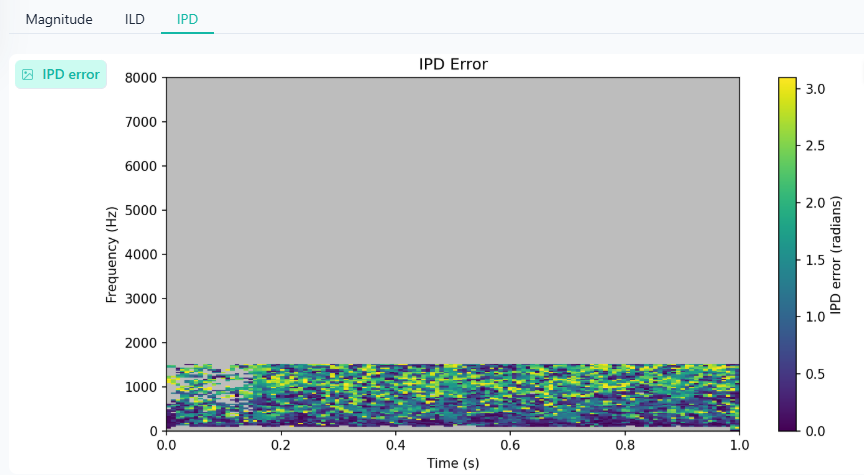}
  \end{minipage}
  \caption{Mode 2 representative example display. After loading a case, the lower panels show the corresponding magnitude, ILD, and IPD error maps together with the audio and metric summaries.}
  \label{fig:app_mode2_maps}
\end{figure}

\subsection{Mode 3: Scene-Level 3D Spatial Error Exploration}
\label{app:mode3}

The third mode provides a scene-level view of error distribution in geometric space.
The user selects a clip from a particular scene and clicks \texttt{Load Scene}.
The visualization displays source and receiver locations inside a 3D box that represents their real spatial relationship in the scene.
The gray ball denotes the sound source.
Other balls denote evaluated target viewpoints, and their colors encode the selected error metric.
The user can switch the \texttt{Metric} control to inspect different failure types, such as magnitude, ILD, IPD, temporal misalignment, or high-frequency error.
Clicking a ball automatically plays the ground-truth and predicted audio for that viewpoint.
This interaction makes it possible to examine whether larger errors are associated with geometric patterns, such as distance, orientation, occlusion, or particular spatial regions of the scene.

\begin{figure}[p]
  \centering
  \includegraphics[width=0.88\linewidth]{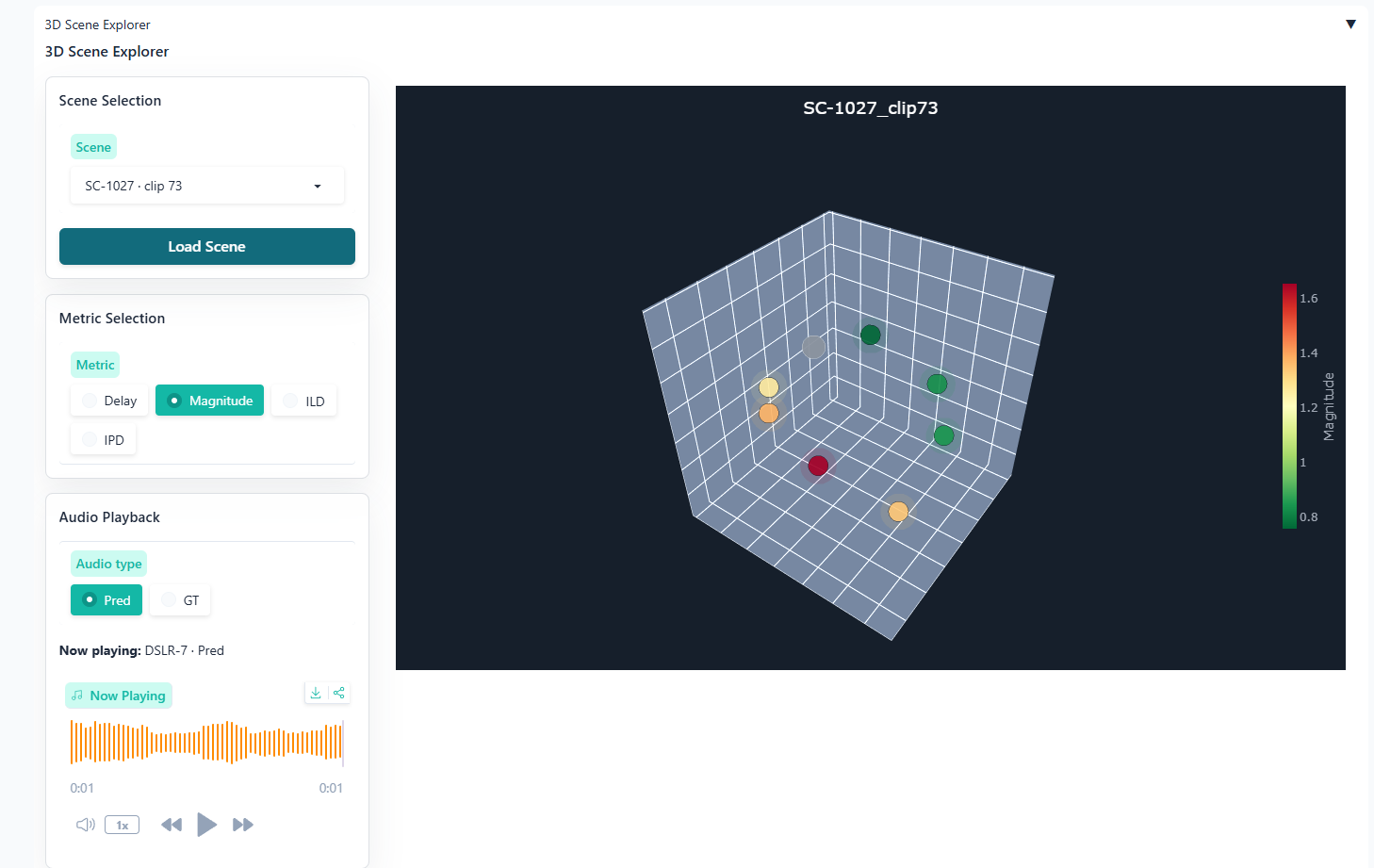}
  \caption{Mode 3 scene-level 3D spatial error view. Ball locations correspond to real geometric relationships in the scene, and color encodes the selected error metric.}
  \label{fig:app_mode3_scene_a}
\end{figure}

\begin{figure}[p]
  \centering
  \includegraphics[width=0.88\linewidth]{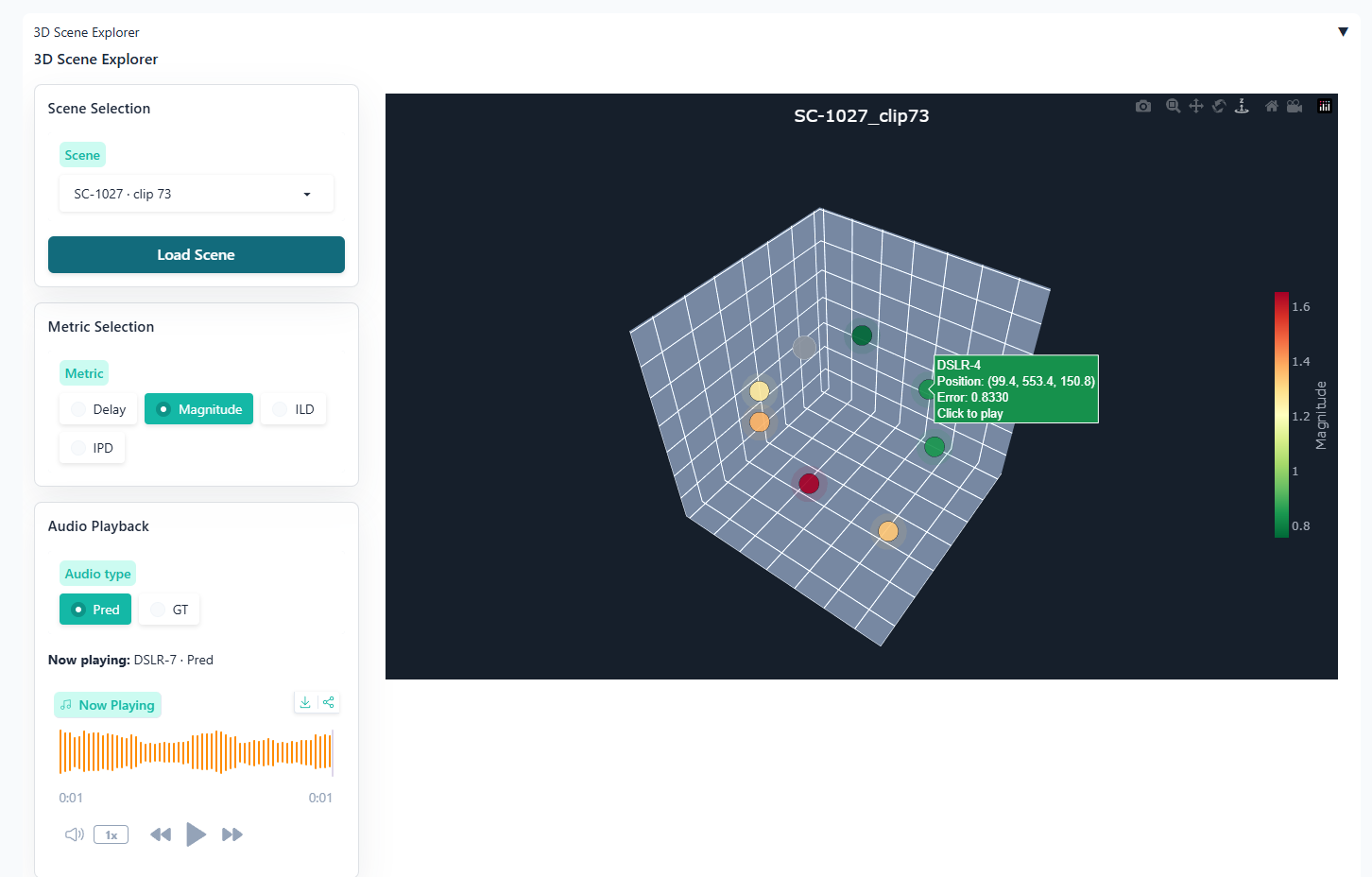}
  \caption{Mode 3 interactive scene analysis. Switching the metric changes the spatial error coloring, and clicking a target viewpoint plays the corresponding ground-truth and predicted audio.}
  \label{fig:app_mode3_scene_b}
\end{figure}

\clearpage

\end{document}